\documentclass[aps,prc,groupedaddress,showpacs]{revtex4}
\usepackage{color,graphicx,bm}
\usepackage{amsmath,bm}

\DeclareMathOperator{\nrd}{\overleftrightarrow{\nabla}}
\DeclareMathOperator{\pard}{\overleftrightarrow{\partial}}
\DeclareMathOperator{\nr}{\overrightarrow{\nabla}}

\begin{document}

\title{Unitary ambiguity of $NN$ contact interactions and the $3N$ force}
\author{L.\ Girlanda$^{\,{\rm a}}$,  A.\ Kievsky$^{\,{\rm b}}$, L.\ E.\ Marcucci$^{\,{\rm b,c}}$, and M.\ Viviani$^{\,{\rm b}}$}
\affiliation{
$^{\,{\rm a}}$\mbox{Dipartimento di Matematica e Fisica, Universit\`a del Salento, and INFN Sezione di Lecce,}\\ \mbox{Via Arnesano, I-73100 Lecce, Italy}\\
$^{\,{\rm b}}$\mbox{INFN, Sezione di Pisa, Largo Bruno Pontecorvo 3, I-56127 Pisa, Italy}\\
$^{\,{\rm c}}$\mbox{Dipartimento di Fisica, Universit\`a di Pisa, Largo Bruno Pontecorvo 3, I-56127 Pisa, Italy}\\
}

\date{\today}

\begin{abstract}
  We identify a redundancy between two- and three-nucleon contact interactions at the fourth and fifth order of the chiral expansion respectively.  In particular we show that tensor-type and spin-orbit three-nucleon contact interactions effectively account for that part of the two-nucleon interaction which depends on the total center-of-mass momentum and is unconstrained by relativity. This might give the chiral effective field theory enough flexibility to successfully address $A=3$ scattering observables already at N3LO.
  \end{abstract}

\pacs{12.39.Fe, 21.30.Fe, 21.45.-v, 21.45.Ff}

\maketitle

\section{Introduction}
The modern understanding of nuclear interactions is based on the chiral effective field theory (ChEFT) framework \cite{vanKolck:1999mw,Bedaque:2002mn,Epelbaum:2005pn,Epelbaum:2008ga,Machleidt:2011zz}. Compared to more phenomenological approaches, a low-momenta power counting allows in principle to improve systematically the accuracy of the theoretical description, pursuing the perturbative expansion to higher and higher orders, and at the same time to assess the theoretical uncertainty introduced by the truncation of the series \cite{Furnstahl:2014xsa,Epelbaum:2014efa}. This is made possible by the approximate chiral symmetry of the underlying quantum chromodynamics (QCD), whose dynamical breakdown is responsible for the emergence of pseudo-Goldstone bosons, the pions, that interact weakly at low energy and are much lighter than all other hadrons.
Pion exchanges among nucleons determine the longest range component of the nuclear interaction, while the dynamics at shorter distances, unresolved by the effective theory, is described in terms of multi-nucleons contact interactions. The associated low-energy constants (LECs), being unconstrained by chiral symmetry, need to be determined from experimental data.  Their number increases as the perturbative series is pushed to higher orders, but their impact should decrease, provided the expansion is well behaved. Obviously, in order to fully exploit the predictive power of the theory, and to put it to a more stringent test, it is important to identify a minimal set of such LECs.

In this paper we concentrate on a redundancy between two nucleon ($NN$) contact couplings which arise at the fourth order of the low-energy expansion (N3LO) and the  subleading three-nucleon ($3N$) contact interactions, which were classified in Ref.~\cite{Girlanda:2011fh} as consisting of 13 independent operators. The  latter  arise  at the fifth order (N4LO) in the ChEFT \cite{Weinberg:1990rz,Weinberg:1991um,Ordonez:1992xp,Ordonez:1993tn,vanKolck:1994yi,Ordonez:1995rz,Epelbaum:2002vt,Bernard:2007sp,Bernard:2011zr}, and as such they should be considered in conjunction with recent accurate versions of the $NN$ interaction developed at N4LO and beyond \cite{Epelbaum:2014efa,Epelbaum:2014sza,Reinert:2017usi,Entem:2017gor,Piarulli:2014bda}. The relevance of these operators has been repeatedly highlighted, in particular for solving long-standing discrepancies in low-energy $N-d$ elastic scattering, like the well known $A_y$ puzzle \cite{Girlanda:2016bze,Girlanda:2016neb,Girlanda:2018xrw}. As we are going to show, five of these operators are equivalent to a suitable  redefinition of the short-range $NN$ potential, realized by specific unitary transformations of the nuclear Hamiltonian.
Such  unitary ambiguities are a common feature in all reductions from a quantum-field theoretical Lagrangian to quantum mechanics and affect also  pion-mediated interactions \cite{Friar:1977xh,Friar:1979by,Pastore:2011ip}. They are systematically exploited in the unitary transformation approach to nuclear forces and electroweak currents \cite{Epelbaum:1998ka,Epelbaum:1999dj} to enforce the renormalizability of nuclear potentials and transition operators \cite{Kolling:2011mt,Krebs:2016rqz}.
We single out in particular two additional transformations that can be used to drop total momentum (${\bf P}$)-dependent $NN$ interactions. Such interactions,  which vanish in the center of mass frame cannot be determined from $NN$ scattering data alone. However, dropping these terms induces {  $3N$ contact interactions}  mostly of tensor and spin-orbit type .

The paper is organized as follows. In section \ref{sec:class} we classify the most general  $NN$ contact unitary transformation at the $O(p^2)$ level. In Section~\ref{sec:rel}  we relate two of these transformations to the  ${\bf P}$-dependent component  of the $NN$ interaction and show that the latter depends on two extra LECs, unconstrained by relativity. In section~\ref{sec:impact} we study the impact of this transformation at the $3N$ level and obtain a reduced form of the subleading $3N$ contact interaction.  Finally, the consequences of the above findings on the structure and convergence of the chiral expansion for $3N$ observables are discussed in section~\ref{sec:concl}.

\section{$NN$ contact unitary transformations at $O(p^2)$ \label{sec:class}}
Following Ref.~\cite{Reinert:2017usi} we write the most general unitary transformation as $U=\exp(\sum_n \alpha_n T_n)$ with $\alpha_n$ real parameters and $T_n$ a complete set of antihermitian operators respecting all underlying symmetries. Since we are interested in purely nucleonic interactions, the generators $T_n$ will only involve local products of nucleon fields,  ordered in the low-energy expansion, according to the number of gradients. The first non trivial case will consists of two-nucleon operators. Rotational, isospin, parity and time-reversal symmetry require the presence of at least two gradients. At this level, a complete  set consists of the following operators:
{\allowdisplaybreaks
\begin{eqnarray}
  T_1 &=& \int d^3{\bf x} N^\dagger \nrd^i N \nabla^i (N^\dagger N)  \sim {\bf k}\cdot {\bf Q}, \\
 % T_{1'} &=& \int d^3{\bf x} N^\dagger \nrd^i \tau^a N \nabla^i  (N^\dagger \tau^a N), \\
  T_2 &=& \int d^3{\bf x} N^\dagger \nrd^i  \sigma^j N \nabla^i  (N^\dagger \sigma^j N)  \sim {\bf k}\cdot {\bf Q} {\bm \sigma}_1 \cdot  {\bm \sigma}_2, \\
 % T_{2'} &=& \int d^3{\bf x} N^\dagger \nrd^i \tau^a \sigma^j N \nabla^i (  N^\dagger \tau^a \sigma^j N), \\
  T_3 &=& \int d^3{\bf x} \left[ N^\dagger \nrd^i  \sigma^i N \nabla^j  (N^\dagger \sigma^j N) + N^\dagger \nrd^i  \sigma^j N \nabla^j  (N^\dagger \sigma^i N) \right]
  \sim {\bf k}\cdot  {\bm \sigma}_1 {\bf Q} \cdot  {\bm \sigma}_2+ {\bf k}\cdot  {\bm \sigma}_2 {\bf Q} \cdot  {\bm \sigma}_1, \\
 % T_{3'} &=& \int d^3{\bf x} \left[ N^\dagger \nrd^i  \tau^a \sigma^i N \nabla^j ( N^\dagger \tau^a \sigma^j N) + N^\dagger \nrd^i  \tau^a \sigma^j N \nabla^j  (N^\dagger \tau^a \sigma^i N) \right], \\
 T_4 &=&  i \epsilon^{ijk} \int d^3{\bf x} N^\dagger \nrd^i N  N^\dagger \nrd^j \sigma^k N  \sim i {\bf P}\times {\bf Q} \cdot ({\bm \sigma}_1 - {\bm \sigma}_2), \\   
 % T_{4'} &=&  i \epsilon^{ijk} \int d^3{\bf x} N^\dagger \tau^a \nrd^i N  N^\dagger \nrd^j \tau^a \sigma^k N, \\   
 T_5 &=& \int d^3{\bf x} \left[ N^\dagger \nrd^i  \sigma^i N \nabla^j ( N^\dagger \sigma^j N) - N^\dagger \nrd^i  \sigma^j N \nabla^j ( N^\dagger \sigma^i N )\right]
 \sim ({\bf P}\cdot {\bm \sigma_1} {\bf k}\cdot {\bm \sigma}_2 -{\bf P}\cdot {\bm \sigma_2} {\bf k}\cdot {\bm \sigma}_1)/2
 , 
  % T_{5'} &=& \int d^3{\bf x} \left[ N^\dagger \nrd^i  \tau^a \sigma^i N \nabla^j  (N^\dagger \tau^a \sigma^j N )- N^\dagger \nrd^i  \tau^a \sigma^j N \nabla^j  (N^\dagger \tau^a \sigma^i N) \right],
\end{eqnarray}
}

\noindent 
where $N^\dagger \nrd^i N = N^\dagger (\nabla^i N) -(\nabla^i N^\dagger ) N$, and $N(x)$ denotes the non-relativistic nucleon field operator.  We have also introduced the dependence on the initial and final relative momenta ${\bf p}$ and ${\bf p'}$, or ${\bf k}={\bf p'} - {\bf p}$ and ${\bf Q}=({\bf p}+{\bf p'})/2$, and on the total momentum ${\bf P}={\bf p}_1 + {\bf p_2}$ of a two-nucleon system. 
The last two generators, which were not considered in Ref.~\cite{Reinert:2017usi}, vanish in the two-nucleon center of mass frame. Their relevance will be clear in the following.
In addition we can also define the corresponding isospin-dependent $T_{n'}$ operators, involving ${\bm \tau}_1\cdot {\bm \tau}_2$, but using the anticommuting nature of nucleon fields and Fierz reshuffling of spin and isospin indeces, one can express them in the { chosen} basis,
\begin{eqnarray}
  T_{1'}&=& -2 T_1 - T_2,\\
  T_{2'}&=&-3 T_1,\\
  T_{3'}&=&-2 T_1 + 2 T_2 -3 T_3,\\
  T_{4'}&=& -2 T_5 - T_4,\\
  T_{5'}&=& -2 T_4 - T_5.
  \end{eqnarray}
When transforming a nuclear Hamiltonian $H$ by the above unitary transformation, one gets additional interactions,
\begin{equation} \label{eq:htransf}
  H \to  U^\dagger H U = H + \sum_n \alpha_n [H,T_n] + ... \equiv H + \sum_n \alpha_n \delta_n H + ... ,
\end{equation}
that amount to a shift of existing LECs, since $H$ already contains all possible interactions allowed by the assumed symmetries. Thus, from the one-body kinetic energy,
\begin{equation} \label{eq:kinetic}
  H_0 = -\frac{1}{2 m} \int d^3{\bf x} N^\dagger \nabla^2 N,
\end{equation}
one gets e.g., using the canonical anticommutation relations \footnote{It is convenient to use the identity $[AB,C]= A \{B,C \} -\{A,C\}B$.}
\begin{equation}
  \delta_1 H_0 = \frac{1}{2 m} \int d^3 {\bf x} \left[ \nabla^i ( N^\dagger \nrd^i \nrd^j N) \nabla^j (N^\dagger N) -  \nabla^i ( N^\dagger \nrd^i  N) \nabla^j (N^\dagger \nrd^j N) \right].
\end{equation}
In the two-nucleon system, the above operator yields an off-shell contribution $\sim ({\bf p}^2- {\bf p'}^2)^2$. In Ref.~\cite{Reinert:2017usi} the unitary  transformations corresponding to $n=1,2,3,$ were used to absorb three of the $O(p^4)$ $NN$ couplings, reducing their number to twelve. As will be shown in Section~\ref{sec:impact}, this also implies the appearance of induced subleading $3N$ contact interactions, which can be written as combinations of the 13 operators introduced in Ref.~\cite{Girlanda:2011fh}. The remaining transformations, corresponding to $n=4,5$,  generate total momentum-dependent interactions which vanish in the center of mass frame, e.g. $\delta_4 H_0 \sim i ({\bf p}^2 - {\bf p'}^2) \,{\bf P} \times ({\bf p} + {\bf p'})\cdot ({\bm \sigma}_1 - {\bm \sigma}_2)$, and will be discussed in the next Section.

\section{\label{sec:rel} ${\bf P}$-dependent $NN$ contact interactions}
Total momentum-dependent interactions are strongly constrained by Poincar\'e symmetry. In the instant form of relativistic dynamics \cite{Dirac:1949cp} the three momentum and angular momentum are the same as in the free theory, described by the Dirac Lagrangian starting from the energy-momentum tensor $T^{\mu\nu}=(i/2) \bar \psi \gamma^\mu \pard^\nu \psi$,  while the Hamiltonian $H$ and boost generators ${\bf K}$ contain the interactions,
\begin{equation}
  H=H_0 + H_I,\quad {\bf K}={\bf K}_0 + {\bf K}_I.
\end{equation}
Since the free generators already satisfy the Poincar\'e commutation relations, the interaction terms must satisfy
\begin{align}
  &  \left[ J^i,K_I^j \right]=i \epsilon^{ijk} K_I^k,\\
&  \left[ K_I^i,P^j \right] =i \delta^{ij} H_I, \label{eq:poincare2}\\
 &  \left[ K_I^i,H_0\right]  + \left[ K^i_0, H_I \right] + \left[ K^i_I,H_I\right] =0,\label{eq:poincare3}\\
 & \left[ K_0^i,K_I^j\right] + \frac{1}{2} \left[ K_I^i,K_I^j \right] - i\leftrightarrow j=0 \label{eq:poincare4}.
\end{align}
The first relation qualifies the interacting boost generator as a vector. The remaining ones are less trivial to satisfy.
In the low-energy theory a non-relativistic reduction can be used to express these operators in terms of the non-relativistic nucleon field $N(x)$ as a series containing increasing powers of soft momenta. For example, the free Hamiltonian and boost generators are expanded as,
\begin{equation}
  H_0=H_0^{(0)} + H_0^{(2)} + H_0^{(4)} + ..., \quad {\bf K}_0 = {\bf K}_0^{(-1)} + {\bf K}_0^{(1)} + {\bf K}_0^{(3)} + ...,
\end{equation}
where the superscripts denote the assigned ``soft power''.
Explicitly,
\begin{align}
  \begin{aligned}
   H_0^{(0)}&=m \int d^3{\bf x} N^\dagger N,  &&H_0^{(2)}=-\frac{1}{8m} \int d^3{\bf x} N^\dagger \nrd^2N, ...\\
   {\bf K}_0^{(-1)}&=m \int d^3{\bf x}\, {\bf x} N^\dagger N,  &&{\bf K}_0^{(1)}=-\frac{1}{8m} \int d^3{\bf x} \,{\bf x} \left[ N^\dagger \nrd^2 N + i \vec{\nabla}\cdot N^\dagger \vec{\sigma}\times \nrd N\right] , ... \, .\\
\end{aligned}
  \end{align}
Contact interactions in $H_I$ can be classified according to the number of participating nucleons,
\begin{equation}
  H_I=H_{NN}+ H_{3N}+ ...,
\end{equation}
and each component can be ordered by the same criterium as
\begin{align}
  H_{NN}&=H_{NN}^{(3)} + H_{NN}^{(5)} + H_{NN}^{(7)} + ... ,\\
  H_{3N}&= H_{3N}^{(6)}+ H_{3N}^{(8)} + ....
\end{align}
The first term in $H_{NN}$ contains the two momentum-independent interactions parametrized by the LECs $C_S$ and $C_T$,
\begin{equation} \label{eq:2ncontact}
  H_{NN}^{(3)}={ \frac{1}{2}}  C_S H_S + { \frac{1}{2}} C_T H_T \equiv  \int d^3{\bf x}\left [ { \frac{1}{2}} C_S N^\dagger N N^\dagger N + { \frac{1}{2}} C_T N^\dagger \vec{\sigma} N \cdot N^\dagger \vec{\sigma} N \right].
\end{equation}
Starting with the following order we can have ${\bf P}$-indepedent or ${\bf P}$-dependent interactions. In $H_{NN}^{(5)}$, the  formers  are parametrized by the LECs $C_{1,...,7}$, while the latters   are unambiguously fixed in terms of the leading LECs $C_S$ and $C_T$ as relativistic $1/m$ corrections \cite{Girlanda:2010ya}. At the following order, $H_{NN}^{(7)}$ contains ${\bf P}$-independent interactions depending on the LECs $D_{1,...,15}$ and a set of ${\bf P}$-dependent ones which have not yet been considered in the literature. Most  of them take the form of relativistic corrections to lower order interactions,  and as such they are fixed unambiguously in terms of the lower-order LECs. Instead, we will be concerned { with} those ${\bf P}$-dependent contributions to $H_{NN}^{(7)}$ which are unconstrained by relativity and depend on extra LECs.
As for the $3N$ interactions, their low-energy expansion starts with a momentum independent term, $H_{3N}^{(6)}$ parametrized by the LEC $c_E$ \cite{Epelbaum:2002vt},
\begin{equation} \label{eq:3ncontact}
  H_{3N}^{(6)}=\frac{c_E}{2 F_\pi^4 \Lambda_\chi}  \int d^3{\bf x} N^\dagger N N^\dagger \tau^a N  N^\dagger \tau^a N ,
\end{equation}
with the pion decay constant $F_\pi$ and the chiral symmetry breaking scale $\Lambda_\chi$ meant to provide the correct scaling based on naive dimensional analysis \cite{Manohar:1983md},
and proceeds with the two-derivatives contact interactions parametrized by the LECs $E_{1,...,13}$ introduced in Ref.~\cite{Girlanda:2011fh},
\begin{equation} \label{eq:contact3np2}
  H_{3N}^{(8)}= \int d^3{\bf x} \sum_{i=1}^{13} E_i  O_i,
\end{equation}
where the explicit expressions for the operators $O_i$ can be read from Eqs.~(14) and (16) of Ref.~\cite{Girlanda:2011fh}.

In order to satisfy the relation~(\ref{eq:poincare2}) a given interaction in $H_I$ implies a corresponding term in ${\bf K}_I$ which we denote as ${\bf W}$, such that
\begin{equation}
  H_I = \int d^3{\bf x} {\cal H}_I(x) \Longrightarrow {\bf W}=\int d^3 {\bf x} \, {\bf x} {\cal H}_I(x).
\end{equation}
The most general form of the interacting part of the boost generator can be written as ${\bf K}_I={\bf W} + \delta {\bf W}$, { i.e. as the sum of the ``minimal'' boost ${\bf W}$ and of an additional term $\delta {\bf W}$ which is translationally invariant ($[P^i,\delta W^j]=0$) and will be denoted as ``intrinsic'', since it is independent of the interacting Hamiltonian. Its  low-energy expansion starts with $\delta {\bf W}^{(4)}$.  At this order} one can list three such operators,
\begin{eqnarray}
  \delta W^i_1 &=& \int d^3 {\bf x} \nr \cdot \left( N^\dagger \vec{\sigma} N\right) N^\dagger \sigma^i N,\\
  \delta W^i_2 &=& i \epsilon^{ijk} \int d^3 {\bf x} \left[  N^\dagger \nrd^j \sigma^k N N^\dagger  N +  N^\dagger \sigma^j N  N^\dagger \nrd^k  N \right],\\
    \delta W^i_3 &=& i \epsilon^{ijk}  \int d^3 {\bf x} \left[  N^\dagger \nrd^i  \sigma^k N N^\dagger  N -  N^\dagger \sigma^j N   N^\dagger \nrd^k  N \right],
\end{eqnarray}
since the operators involving ${\bm \tau}_1\cdot {\bm \tau}_2$ are Fierz related to the ones above.
These intrinsic boosts were ignored in Ref.~\cite{Girlanda:2010ya}, since they do not play a role at the order considered there, respectively $O(p^4)$ and $O(p^3)$ for the relations~(\ref{eq:poincare3}) and (\ref{eq:poincare4}),
\begin{align}
  &\left[ {\bf K}_0^{(-1)},H_I^{(5)}\right] + \left[ {\bf K}_0^{(1)},H_I^{(3)}\right]
  + \left[ {\bf W}^{(4)},H_0^{(0)}\right]  + \left[ \delta {\bf W}^{(4)},H_0^{(0)}\right]   + \left[ {\bf W}^{(2)},H_0^{(2)}\right] =0, \label{eq:poincarep4}\\
  &    \left[ K_0^{ (-1)i}, W^{ (4)j} \right] +  \left[ K_0^{ (1)i}, W^{ (2)j} \right]
  +   \left[ K_0^{ (-1)i}, \delta W^{ (4)j} \right] - i \leftrightarrow j =0. \label{eq:poincarep3}
\end{align}
Indeed, in the first of the above equations, $\delta {\bf W}$ is irrelevant, since it commutes with $H_0^{(0)}$. Moreover, as found in Ref.~\cite{Girlanda:2010ya}, Eq.~(\ref{eq:poincarep3}) without the commutators involving $\delta {\bf W}$ is valid as a consequence of Eq.~(\ref{eq:poincarep4}). This means that we must have
\begin{equation}
  [K_0^{ (-1)i},\delta W^{ (4)j}] - i\leftrightarrow j =0,
\end{equation}
which rules out $\delta {\bf W}_3$.  In other words, only the ${\bf P}$-independent intrinsic boosts $\delta {\bf W}_1$ and $\delta {\bf W}_2$ are allowed, and we can write the most general intrinsic boost $ \delta {\bf W}^{(4)}$ in terms of two constants, $ \delta {\bf W}^{(4)} =\sum_{i=1}^2 \beta_i \delta {\bf W}_i$.  The two independent intrinsic boost generators are related to the transformations $T_4$ and $T_5$ of the previous section by the following relations,
\begin{equation} \label{eq:unitboost}
  \left[ {\bf K}_0^{(-1)}, T_5 \right] = -4 \delta {\bf W}_1, \quad   \left[ {\bf K}_0^{(-1)}, T_4 \right] = -2 \delta {\bf W}_2.
\end{equation}
They start to play a role at the orders $O(p^6)$ and $O(p^5)$ respectively,
\begin{align} 
  0  &=\left[ {\bf W}^{(6)}, H_0^{(0)} \right] + \left[ {\bf W}^{(4)}, H_0^{(2)} \right] +  \left[{\bf K}_0^{(1)},H_I^{(5)}\right]  +  \left[{\bf W}^{(2)}, H_0^{(4)}\right] + \left[{\bf K}_0^{(3)},H_I^{(3)}\right] \nonumber \\
&
 + \left[\delta {\bf W}^{(4)}, H_0^{(2)}\right] +  \left[ {\bf K}_0^{(-1)}, H_I^{(7)}\right] ,\label{eq:poincarep6}\\
0&= \left[ K_0^{ (-1)i}, W^{ (6)j} \right] + \left[K_0^{ (1)i},W^{(4)j}\right] + \left[K_0^{ (1)i},\delta W^{(4)j}\right]  +  \left[K_0^{ (3)i},W^{(2)j}\right] - i\leftrightarrow j\label{eq:poincarep5},
\end{align}
which again involve only two-nucleon terms. From these equations the relativistic corrections in $H_I^{(7)}$ can be identified, following the steps of Ref.~\cite{Girlanda:2010ya}.
 Indeed,
the  interactions in $H_I^{(7)}$, with corresponding minimal boosts ${\bf W}^{(6)}$, must satisfy the above constraints. Considering Eq.~(\ref{eq:poincarep6}), the first term vanishes, the second and third terms represent $1/m$ corrections to the interactions involving $C_i$, the fourth and fifth terms represent $1/m^3$ corrections to the interactions involving $C_S$ and $C_T$. Ignoring these $1/m$ corrections we are left either with ${\bf P}$-independent terms in $H_I^{(7)}$, which commute with ${\bf K}_0^{(-1)}$, and thus satisfy Eq.~(\ref{eq:poincarep6}) with $\delta {\bf W}^{(4)}=0$ (these are the operators multiplied by $D_1$,...,$D_{15}$)  or with  ${\bf P}$-dependent terms whose commutator with ${\bf K}_0^{(-1)}$ must be compensated by the terms involving $\delta {\bf W}^{(4)}$.
  Thus there is a one-to-one correspondence between the possible forms of the intrinsic boost $\delta {\bf W}^{(4)}$ (which we already classified as $\delta {\bf W}_1$ and $\delta {\bf W}_2$)  and the allowed ${\bf P}$-dependent interactions of $H_I^{(7)}$. It is possible to show that Eq.~(\ref{eq:poincarep5}) then follows from Bianchi identities.

  In view of  Eqs.~(\ref{eq:unitboost}), the above statement can be immediately understood:
 a unitary transformation involving $T_4$ and $T_5$ will generate from $H_0$ some ${\bf P}$-dependent interaction terms of $H_I^{(7)}$. At the same time, as it is clear from Eq.~(\ref{eq:unitboost}), from the free boost generator ${\bf K}_0$ one gets the interacting intrinsic boosts that exactly compensate for these terms, such that the Poincar\'e commutation relations (\ref{eq:poincarep6}) and (\ref{eq:poincarep5}) remain satisfied,  as they should by unitarity.
This means that there are two ${\bf P}$-dependent $NN$ contact interactions  in $H_I^{(7)}$ which are completely unconstrained, depending on two free  LECs.
A possible parametrization of the resulting ${\bf P}$-dependent $NN$ potential is in terms of two extra LECs, $D_{16}$ and $D_{17}$, such that
  {
\begin{align} \label{eq:pdepvnn}
  V_{NN}({\bf P} ) &= i D_{16} {\bf k}\cdot {\bf Q} \, {\bf Q}\times {\bf P}\cdot ({\bm \sigma}_1 - {\bm \sigma}_2) + D_{17} {\bf k}\cdot {\bf Q}\, ({\bf k}\times {\bf P})\cdot ({\bm \sigma}_1 \times {\bm \sigma}_2) \nonumber \\
  &=   i \left[ (D_{16} -D_{17}) + D_{17} {\bm \tau}_1 \cdot {\bm \tau}_2 \right]  {\bf k} \cdot {\bf Q} \, {\bf Q} \times {\bf P} \cdot ({\bm \sigma}_1 - {\bm \sigma}_2), 
\end{align}
where the last equality follows from the Fierz identities.
These interactions change the spin of the $NN$ system and they are both linear in ${\bf P}$. They act differently in the two isospin channels \footnote{Notice that, away from the center-of-mass system, the usual relation $(-1)^{L+S+T}=-1$ between the orbital angular momentum $L$, the spin $S$ and the isospin $T$ of the $NN$ pair is not necessarily satisfied, due to the presence of the angular momentum associated to the overall motion.}.

The ${\bf P}$-dependent potential (\ref{eq:pdepvnn}) bears some resemblance with the Thomas precession term \cite{Forest:1995sg,Krajcik:1974nv,Friar:1975zza} which is a relativistic ${\bf P}$-dependent interaction determined by the center-of-mass potential $v$ as
\begin{equation} \label{eq:thomas}
\delta  v_{\mathrm{TP}}= \frac{i}{8 m^2} \left[ ({\bm \sigma}_1 - {\bm \sigma}_2)\times {\bf P} \cdot {\bf p}, v \right],
\end{equation}
and is part of the terms discussed in the previous paragraph. We emphasize once more that the ${\bf P}$-dependent interaction (\ref{eq:pdepvnn}) is of a different nature, and the corresponding LECs are not fixed as $1/m$ corrections to lower order interactions. For example, { if we take for $v$ in Eq.~(\ref{eq:thomas}) the subleading two-nucleon contact potential, defined as customary}
\begin{align}
  v &=C_1 k^2 + C_2 k^2 {\bm \tau}_1 \cdot {\bm \tau}_2 + C_3 k^2 {\bm \sigma}_1 \cdot {\bm \sigma}_2 + C_4 k^2 {\bm \sigma}_1 \cdot {\bm \sigma}_2 \, {\bm \tau}_1 \cdot {\bm \tau}_2 \nonumber \\
  & + C_5 S_{12}({\bf k}) + C_6 S_{12}({\bf k})  {\bm \tau}_1 \cdot {\bm \tau}_2 + i C_7 {\bf S}\cdot {\bf Q} \times {\bf k},
\end{align}
with $S_{12}({\bf k})=3 {\bm \sigma}_1 \cdot {\bf k}\, {\bm \sigma}_2 \cdot {\bf k} - k^2 {\bm \sigma}_1 \cdot {\bm \sigma}_2$ and ${\bf S}=({\bm \sigma}_1 + {\bm \sigma}_2)/2$,
then the induced Thomas precession potential   is
\begin{align}
  \delta v_{\mathrm{TP}} &= \frac{i}{8 m^2} {\bf P} \times {\bf k} \cdot ({\bm \sigma}_1 - {\bm \sigma}_2 ) \left\{ k^2 \left[ C_1 - C_3 + C_5 + \left( C_2 - C_4 + C_6 \right) {\bm \tau}_1\cdot {\bm \tau}_2 \right] \right. \nonumber \\
  &\left. -4 Q^2 \left[ C_3 + 3 C_4 - C_5 - 3 C_6  + \left( C_3 - C_4 - C_5 + C_6 \right) {\bm \tau}_1 \cdot {\bm \tau}_2 \right] \right\} \nonumber \\
  &+ \frac{3 i}{4 m^2} \left[ {\bf P}\cdot {\bf Q}\, {\bf k}\times {\bf Q}\cdot ({\bm \sigma}_1 - {\bm \sigma}_2 ) - {\bf k}\cdot {\bf Q} \,{\bf P}\times {\bf Q}\cdot ({\bm \sigma}_1 - {\bm \sigma}_2 ) \right] \left[ C_5 + 3 C_6 + \left(C_5 - C_6\right) {\bm \tau}_1 \cdot {\bm \tau}_2 \right] \nonumber \\
  &  - \frac{i C_7}{16 m^2} {\bf P}\times {\bf Q} \cdot {\bf k} \,{\bf Q}\cdot ({\bm \sigma}_1 - {\bm \sigma}_2 ) \left( 1 - {\bm \tau}_1 \cdot {\bm \tau}_2 \right).
\end{align}

Thus we see that part of the interactions in Eq~(\ref{eq:pdepvnn}) are indeed also generated by the Thomas precession, with fixed coefficients, while the LECs $D_{16}$ and $D_{17}$ are completely unconstrained. 
They} cannot be determined from $NN$ scattering data, but only in $A>2$ systems, or as a high-order contribution to the two-nucleon electromagnetic current.  Their contribution vanishes on shell and, similarly to the 3 combinations of the $D_i$ already identified in Ref.~\cite{Reinert:2017usi}, they can be absorbed by a unitary transformation. However, in so doing  one obtains at the same time an induced $3N$ interaction, as will be discussed in the next Section.

\section{Impact on the $3N$ sector \label{sec:impact}}
If $H$ in Eq.~(\ref{eq:htransf}) is a two-nucleon operator then the unitary transformations defined in Section~\ref{sec:class} generate three-nucleon operators. At the leading order we have the following contributions induced by the transformations of $H_S$ and $H_T$ appearing in Eq.~(\ref{eq:2ncontact}),
{\allowdisplaybreaks
  \begin{eqnarray}
  \delta_1 H_S &=& -4 \int d^3{\bf x} \nabla^i (N^\dagger N)  \nabla^i (N^\dagger N) (N^\dagger N),\\
  \delta_1 H_T &=& -4 \int d^3{\bf x} \nabla^i (N^\dagger\sigma^j N) \nabla^i (N^\dagger N) (N^\dagger  \sigma^j N),\\
  \delta_2 H_S &=& \delta_1 H_T,\\
  \delta_2 H_T &=& 4 \int d^3{\bf x} \left[ i \epsilon^{ijk}  \nabla^l (N^\dagger \sigma^i N)   N^\dagger \nrd^l \sigma^j N N^\dagger \sigma^k N - \nabla^j (N^\dagger \sigma^i N)   \nabla^j (N^\dagger \sigma^i N ) N^\dagger  N \right],\\
  \delta_3 H_S &=& -4 \int d^3{\bf x} \left[ \nabla^i ( N^\dagger \sigma^j N) \nabla^j (N^\dagger N) N^\dagger \sigma^i N + \nabla^i ( N^\dagger \sigma^i N) \nabla^j (N^\dagger N)  N^\dagger \sigma^j N\right],\\
  \delta_3 H_T &=& 4 \int d^3{\bf x} \left[i \epsilon^{ijk} \nabla^i ( N^\dagger \sigma^l N)   (N^\dagger \nrd^l \sigma^j N)  N^\dagger \sigma^k N + i \nabla^l  ( N^\dagger \sigma^l N)  N^\dagger \nrd^i  \sigma^j N  N^\dagger \sigma^k N\right. \nonumber \\
  && \left. -\nabla^i ( N^\dagger \sigma^j N) \nabla^j (N^\dagger\sigma^i N) N^\dagger  N  - \nabla^i  ( N^\dagger \sigma^i N) \nabla^j  (N^\dagger\sigma^j N)  N^\dagger   N\right],\\
  \delta_4 H_S &=& 4 i \epsilon^{ijk} \int d^3{\bf x} \left[ \nabla^i (N^\dagger N)  N^\dagger \nrd^j  N  N^\dagger \sigma^k N    -\nabla^i (N^\dagger N)  N^\dagger \nrd^j \sigma^k N N^\dagger N\right] , \\    
  \delta_4 H_T &=& -4  \int d^3{\bf x} \left[ i \epsilon^{ijk} \nabla^i (N^\dagger \sigma^l N)  N^\dagger \nrd^j \sigma^k N  N^\dagger \sigma^l N    + i \epsilon^{ijk} \nabla^i (N^\dagger \sigma^j N)  N^\dagger \nrd^k  N N^\dagger N \right. \nonumber \\
    && \left. + N^\dagger \nrd^i N N^\dagger \nrd^j \sigma^i N N^\dagger \sigma^j N - N^\dagger \nrd^i \sigma^i N N^\dagger \nrd^j N  N^\dagger \sigma^j N \right], \\
  \delta_5 H_S &=& 0, \\
  \delta_5 H_T &=& -4  \int d^3{\bf x} \left[ i \epsilon^{ijk} \nabla^i (N^\dagger \sigma^l N)  N^\dagger \nrd^l \sigma^j N N^\dagger \sigma^k N - \nabla^l  (N^\dagger \sigma^l N) N^\dagger \nrd^i \sigma^j N \cdot N^\dagger \sigma^k N \right. \nonumber \\
    && \left. + \nabla^i (N^\dagger \sigma^i N) \nabla^j  (N^\dagger \sigma^j N) N^\dagger N - \nabla^i (N^\dagger \sigma^j N) \nabla^j (N^\dagger \sigma^i N) N^\dagger N \right].
\end{eqnarray}
}
These operators can be expressed in terms of the basis $\{ O_i \}$ defined in Eq.~(\ref{eq:contact3np2}), using the identities derived in Ref.~\cite{Girlanda:2011fh}.
As a result, the general unitary transformation of the two-nucleon Hamiltonian in Eq.~(\ref{eq:2ncontact}) produces the following contribution,
\begin{equation}
  \left[ H_{NN},\sum_{n=1}^5 \alpha_n T_n \right] =\int d^3{\bf x}  \sum_{i=1}^{13}  \, \delta E_i \, O_i,
\end{equation}
with
{\allowdisplaybreaks
  \begin{eqnarray}
  \delta E_1 &=&  \alpha_1 \left( C_S + C_T \right)+   \alpha_2 \left( C_S - 2 C_T \right), \label{eq:delta1}\\
  \delta E_2 &=&  3 \alpha_2  C_T  +2 \alpha_3 C_T -4  \alpha_4 C_T +2  \alpha_5 C_T,\label{eq:delta2}\\
  \delta E_3 &=& 2 \alpha_1 C_T  +  \alpha_2 \left( 2 C_S -  C_T \right) + \frac{2}{3} \alpha_3 \left( 2 C_S - C_T \right) +4 \alpha_4 C_T- 2  \alpha_5 C_T,\label{eq:delta3}\\
  \delta E_4 &=& \frac{2}{3} \alpha_1 C_T  + \frac{1}{3} \alpha_2 \left( 2 C_S -  7 C_T \right) - \frac{2}{3} \alpha_3  C_T  +\frac{4}{3} \alpha_4 C_T -\frac{2}{3} \alpha_5 C_T,\label{eq:delta4}\\
  \delta E_5 &=& 2 \alpha_1 C_T  + 2 \alpha_2 \left(  C_S - 2 C_T \right) + \frac{2}{3} \alpha_3 \left( 2 C_S - C_T \right) +4 \alpha_4 C_T-2 \alpha_5 C_T,\label{eq:delta5}\\ 
  \delta E_6 &=& \frac{2}{3} \alpha_1 C_T  + \frac{2}{3} \alpha_2 \left(  C_S - 2 C_T \right) -\frac{2}{3} \alpha_3  C_T +\frac{4}{3} \alpha_4 C_T-\frac{2}{3} \alpha_5 C_T,\label{eq:delta6}\\
  \delta E_7 &=& 8 \alpha_4 C_T ,\label{eq:delta7}\\
\delta E_8 &=& \frac{1}{3} \delta E_7,\label{eq:delta8}\\
\delta E_9 &=& 3\alpha_1  C_T +  3 \alpha_2 (C_S-2 C_T) +2 \alpha_3  (C_S -2 C_T) -  \alpha_4 \left(C_S - 5 C_T \right) - 4 \alpha_5 C_T,\label{eq:delta9}\\
\delta E_{10} &=& \alpha_1  C_T  + \alpha_2 \left(C_S - 2 C_T \right)-\frac{1}{3} \alpha_4 \left(3 C_S - 7 C_T \right),\label{eq:delta10} \\
\delta E_{11} &=& 3\alpha_1  C_T  +3 \alpha_2 \left(C_S - 2 C_T \right)+2 \alpha_3 \left( C_S - 2 C_T \right) +   \alpha_4 \left( C_S - 5 C_T \right) + 4 \alpha_5 C_T,\label{eq:delta11}\\
\delta E_{12} &=& \alpha_1  C_T  + \alpha_2 \left(C_S - 2 C_T \right)+  \frac{1}{3}  \alpha_4 \left( 3 C_S - 7 C_T \right) ,\label{eq:delta12}\\
\delta E_{13} &=& -8\alpha_4  C_T  +4 \alpha_5 C_T ,\label{eq:delta13}
\end{eqnarray}
}
which amounts to a shift of the thirteen subleading LECs $E_i$  in Eq.~(\ref{eq:contact3np2}),
\begin{equation} \label{eq:shifts}
  H_{3N}^{(8)}  \to \int d^3 {\bf x} \sum_{i=1}^{13} (E_i + \delta E_i) O_i \equiv \int d^3 {\bf x} \sum_{i=1}^{13} E_i^{(\alpha)} O_i.
\end{equation}
Therefore, for all nonzero values of $C_S$ and $C_T$, the five parameters $\alpha_n$ defining the unitary transformation can be chosen so that five of the thirteen subleading LECs  can be eliminated, i.e. $E^{(\alpha)}_i=0$. The sum in Eq.~(\ref{eq:contact3np2}) can then be restricted e.g. to $i=2,...,9$. This happens at the price of considering a more elaborate $NN$ interaction,  comprising all of the N4LO LECs $D_{1,...,15}$ as well as the ones parametrizing the ${\bf P}$-dependent $NN$ interaction.

Another point of view can be adopted. It is generally accepted that the $3N$ interaction is parameter-free at N3LO \cite{Bernard:2007sp,Bernard:2011zr}, the LECs $E_i$  of Eq.~(\ref{eq:contact3np2}) contributing only at N4LO. 
However, the (N3LO) effect of the ${\bf P}$-dependent, or of the off-shell component of the $NN$ interaction is equivalent to  the above contact $3N$ interaction. Thus the shifts $\delta E_i$ in Eq.~(\ref{eq:shifts})  are to be regarded as an effect at N3LO.  In other words, the five LECs parametrizing the N3LO $NN$ off-shell interaction can be fitted to observables of the $3N$ system and interpreted as a $3N$ interaction.

We emphasize in particular the role of the ${\bf P}$-dependent $NN$ interaction,
 which has never been taken into account, since it cannot be determined from the $NN$ scattering data.
 Even if the complete $NN$ interaction (including all of the  LECs $D_{1,...,15}$) is used in $3N$ calculations, the necessity to discard the ${\bf P}$-dependent terms leads to the appearance of a subleading $3N$ contact interaction already at N3LO. The exact form of this interaction can be read from Eqs.~(\ref{eq:delta1})-(\ref{eq:delta13}) by inspecting the terms proportional to $\alpha_4$ and $\alpha_5$, which can be regarded as a sort of LECs. In other words, supplementing the $NN$ interaction with its ${\bf P}$-dependent component, Eq.~(\ref{eq:pdepvnn}), the corresponding interactions can be absorbed by a unitary transformation with parameters
\begin{equation} \label{eq:alpha45}
  \alpha_4=-\frac{m}{8} D_{16}, \quad \alpha_5=-\frac{m}{4} D_{17},
\end{equation}
i.e. by the transformation of the kinetic energy operator of Eq.~(\ref{eq:kinetic}),
\begin{equation}
[H_0, \alpha_4 T_4 + \alpha_5 T_5]= - V_{NN}({\bf P}),
\end{equation}
that in turn generates the $3N$ couplings $E_i$'s according to Eqs.~(\ref{eq:delta1})-(\ref{eq:delta13}).

  Thus we can say that, contrary to the commonly accepted wisdom, the $3N$ force is not parameter-free at N3LO, but depends on five LECs. Three of them are combinations of the LECs $D_1$, ..., $D_{15}$, if one removes them from the $NN$ potential, as done in Ref.~\cite{Reinert:2017usi}. Two more correspond to the new LECs $D_{16}$ and $D_{17}$, which can be viewed as contributions to the subleading $3N$ contact potential, 
\begin{eqnarray} \label{eq:v3n}
  V_{3N} &=& \frac{m}{8} D_{16} \biggl[ C_S (O_9 + O_{10} - O_{11} - O_{12})  \nonumber \\
    && \left. - C_T \left( 4 O_3 + \frac{4}{3} O_4 + 4 O_5 + \frac{4}{3} O_6 + 8 O_7 + \frac{8}{3} O_8 + 5 O_9 -\frac{7}{3} O_{10} - 5 O_{11} - \frac{7}{3} O_{12} - 8 O_{13} \right) \right] \nonumber \\
  && - \frac{m}{2} D_{17} C_T \left( O_2 - O_3 -\frac{1}{3} O_4 - O_5 - \frac{1}{3} O_6 - 2 O_9 + 2 O_{11} + 2 O_{13}\right).
\end{eqnarray}
Due to the fact that, on phenomenological grounds, $C_S \gg C_T$, the main effect of the new ${\bf P}$-dependent $NN$ interactions amounts in this limit  to the first line of the above equation, and involves a single LEC $m D_{16}$.
On the other hand, the large numerical coefficients multiplying $C_T$  in $\delta E_7$, i.e. most notably for the spin-orbit operator $O_7$, might explain its instrumental role in the resolution of the $A_y$ puzzle of low-energy $N-d$ scattering \cite{Girlanda:2016bze,Girlanda:2016neb,Girlanda:2018xrw,Kievsky:1999nw}.

  Notice that the order mismatch between N3LO and N4LO is removed in the Weinberg counting, i.e. $m\sim O(\Lambda_\chi^2/p)$ \cite{Reinert:2017usi}. On the other hand,
  if one explicitly includes a factor of $1/m$ in the contact LECs  parametrizing the $NN$ potential, in order to give the latter the same scaling as the kinetic energy, then the unitary transformation becomes independent of $m$, as in Eq.~(\ref{eq:alpha45}), corresponding to the possibility of removing the $m$ dependence  from the non-relativistic theory altogether. The potential in Eq.~(\ref{eq:v3n}) would also scale as $1/m$ in this case, and no enhancement would be formally obtained as compared to the leading $3N$ contact interaction, provided a factor of $1/m$ is also attached to $c_E$ in Eq.~(\ref{eq:3ncontact}).

  The same unitary transformation also affects another class of contributions to the $3N$ interaction, with one-pion-exchange/contact topology, of the same type as the so called $c_D$ term \cite{Epelbaum:2002vt}. Indeed, if $H$ in Eq.~(\ref{eq:htransf})  is the $\pi N$ Hamiltonian,
\begin{equation}
  H_{\pi N} = \frac{g_A}{2 F_\pi} \int d^3{\bf x} N^\dagger {\bm \nabla} \pi^a \cdot {\bm \sigma} \tau^a  N,
\end{equation}
the induced Hamiltonian is a $\pi N N$ coupling, which gives rise to a $3N$ potential of the following form,
\begin{align}
  V_{3NF} &= -\frac{g_A}{F_{\pi}} \sum_{i\neq j \neq k} \frac{{\bf k}_k \cdot {\bm \sigma}_k \, {\bm \tau}_i\cdot {\bm \tau}_k}{k_k^2 + m_\pi^2} \left\{ \alpha_1 {\bf k}_j \cdot {\bf k}_k \, {\bf k}_k \cdot {\bm \sigma}_i \right. \nonumber \\
  &+ \alpha_2 \left[ {\bf k}_j \cdot {\bf k}_k \, {\bf k}_k \cdot {\bm \sigma}_j + 2 i {\bf k}_j\cdot ({\bf Q}_i - {\bf Q}_j) \, {\bf k}_k \cdot {\bm \sigma}_i \times {\bm \sigma}_j \right] \nonumber \\
  &+ (\alpha_3 + \alpha_5) \left[ k_k^2  {\bf k}_j \cdot {\bm \sigma}_j - 2 i {\bf k}_j\cdot {\bm \sigma}_j  \, {\bf k}_k \cdot {\bf Q}_i \times {\bm \sigma}_i
+ 2 i {\bf Q}_j \cdot {\bm \sigma}_j \, {\bf k}_k \cdot {\bf k}_j \times {\bm \sigma}_i \right]  \nonumber \\
  &+ (\alpha_3 - \alpha_5) \left[ {\bf k}_j \cdot {\bf k}_k   \, {\bf k}_k \cdot {\bm \sigma}_j + 2 i {\bf k}_j\cdot {\bm \sigma}_j  \, {\bf k}_k \cdot {\bf Q}_j \times {\bm \sigma}_i
- 2 i {\bf Q}_i \cdot {\bm \sigma}_j \, {\bf k}_k \cdot {\bf k}_j \times {\bm \sigma}_i \right]  \nonumber \\
  & \left. -2 \alpha_4 \left[ {\bf k}_k \cdot {\bm \sigma}_i \, {\bf k}_k \cdot {\bf Q}_j \times {\bm \sigma}_j - 2 i {\bf k}_k \cdot {\bf Q}_i \, \left( {\bf k}_k \cdot {\bf Q}_i \, {\bf Q}_j \cdot {\bm \sigma}_i - {\bf k}_k \cdot {\bf Q}_j \, {\bf Q}_i \cdot {\bm \sigma}_i \right) \right] \right\}.
\end{align}
Similarly to the purely contact ones, these contributions, which are nominally N4LO, are enhanced by a factor of $m$, and therefore promoted to N3LO. To the best of our knowledge they were not considered in the literature before. It will be interesting to study their effects in $3N$ scattering observables. 

It is worth mentioning that no genuine three-nucleon unitary transformation can be used to the same purpose. For instance, taking
\begin{equation}
  T_{3N} = i \int d^3{\bf x} N^\dagger \vec{\sigma} N \times N^\dagger\vec{\sigma} N \cdot N^\dagger \vec{\sigma} N,
\end{equation}
then $[H_0,T_{3N}]=0$, due to the antisymmetry with respect to the nucleon labels.

\section{Concluding remarks \label{sec:concl}}
By examining the most general $NN$ contact unitary transformation we have identified the precise relation between the subleading contact $3N$ interaction and specific off-shell or ${\bf P}$-dependent components of the $NN$ interaction.
%The separation between the two- and three-nucleon interactions is a matter of arbitrary choice.
Provided the full $NN$ interaction up to $O(p^4)$ is determined from experimental data, including the off-shell components encoded in the LECs $D_i$ and the ${\bf P}$-dependent contributions (which might require considering electromagnetic observables depending on the same LECs), then five of the $3N$ subleading LECs $E_i$ become redundant. 
Alternatively, one can disregard these contributions in the two-nucleon systems, at the price of a more involved $3N$ contact interaction.
We have identified in particular the two unitary transformations that allow to drop the free LECs parametrizing the ${\bf P}$-dependent component of the $NN$ interaction, Eq.~(\ref{eq:pdepvnn}), which cannot be determined from a fit to $NN$ scattering data alone. These interactions would contribute in larger systems, like $A=3$, together with ``drift terms'' representing relativistic $1/m$ corrections, but they are never considered in actual calculations. Their effect can be traded with  two specific combinations  of the subleading $3N$ contact operators, which can be read from Eqs.~(\ref{eq:delta1})-(\ref{eq:delta13}) as the contributions proportional to $\alpha_4$ and $\alpha_5$, the latter given in turn by Eqs.~(\ref{eq:alpha45}).
We notice that these unitary transformations reshuffle the individual terms of the low-energy expansion. In particular, by absorbing the N3LO $NN$ contact LECs  $D_i$'s, their effect is attributed to the N4LO $3N$ contact LECs $E_i$'s { and give rise additionally to a $3N$ force of one-pion-exchange/contact type.} Therefore, we argue that this procedure could modify the expected convergence pattern of the chiral series, and explain the difficulty of the N3LO $3N$ chiral interactions to address the $A=3$ scattering observables \cite{Golak:2014ksa} justifying the observed prominent role of the  spin-orbit and tensor $3N$ contact interactions \cite{Girlanda:2018xrw}.


\begin{thebibliography}{100}

\bibitem{vanKolck:1999mw}
U.~van Kolck,
%``Effective field theory of nuclear forces,''
Prog.\ Part.\ Nucl.\ Phys.\  \textbf{43}, 337-418 (1999)
doi:10.1016/S0146-6410(99)00097-6
[arXiv:nucl-th/9902015 [nucl-th]].
%255 citations counted in INSPIRE as of 04 Apr 2020

\bibitem{Bedaque:2002mn}
P.~F.~Bedaque and U.~van Kolck,
%``Effective field theory for few nucleon systems,''
Ann.\ Rev.\ Nucl.\ Part.\ Sci.\  \textbf{52}, 339-396 (2002)
doi:10.1146/annurev.nucl.52.050102.090637
[arXiv:nucl-th/0203055 [nucl-th]].
%618 citations counted in INSPIRE as of 04 Apr 2020

%\cite{Epelbaum:2005pn}
\bibitem{Epelbaum:2005pn}
E.~Epelbaum,
%``Few-nucleon forces and systems in chiral effective field theory,''
Prog.\ Part.\ Nucl.\ Phys.\  \textbf{57}, 654-741 (2006)
doi:10.1016/j.ppnp.2005.09.002
[arXiv:nucl-th/0509032 [nucl-th]].
%463 citations counted in INSPIRE as of 04 Apr 2020

%\cite{Epelbaum:2008ga}
\bibitem{Epelbaum:2008ga}
E.~Epelbaum, H.~Hammer and U.~Meissner,
%``Modern Theory of Nuclear Forces,''
Rev.\ Mod.\ Phys.\  \textbf{81}, 1773-1825 (2009)
doi:10.1103/RevModPhys.81.1773
[arXiv:0811.1338 [nucl-th]].
%1112 citations counted in INSPIRE as of 04 Apr 2020

%\cite{Machleidt:2011zz}
\bibitem{Machleidt:2011zz}
R.~Machleidt and D.~Entem,
%``Chiral effective field theory and nuclear forces,''
Phys.\ Rept.\  \textbf{503}, 1-75 (2011)
doi:10.1016/j.physrep.2011.02.001
[arXiv:1105.2919 [nucl-th]].
%860 citations counted in INSPIRE as of 04 Apr 2020


\bibitem{Furnstahl:2014xsa}
R.~Furnstahl, D.~Phillips and S.~Wesolowski,
%``A recipe for EFT uncertainty quantification in nuclear physics,''
J.\ Phys.\ G \textbf{42}, no.3, 034028 (2015)
doi:10.1088/0954-3899/42/3/034028
[arXiv:1407.0657 [nucl-th]].
%76 citations counted in INSPIRE as of 04 Apr 2020

\bibitem{Epelbaum:2014efa}
E.~Epelbaum, H.~Krebs and U.~Meissner,
%``Improved chiral nucleon-nucleon potential up to next-to-next-to-next-to-leading order,''
Eur.\ Phys.\ J.\ A \textbf{51}, no.5, 53 (2015)
doi:10.1140/epja/i2015-15053-8
[arXiv:1412.0142 [nucl-th]].
%265 citations counted in INSPIRE as of 04 Apr 2020

%\cite{Girlanda:2011fh}
\bibitem{Girlanda:2011fh}
L.~Girlanda, A.~Kievsky and M.~Viviani,
%``Subleading contributions to the three-nucleon contact interaction,''
Phys.\ Rev.\ C \textbf{84}, 014001 (2011)
doi:10.1103/PhysRevC.84.014001. Erratum  Phys.\ Rev.\ C \textbf{102}, 019903 (2020) doi:10.1103/PhysRevC.84.014001 [arXiv:1102.4799 [nucl-th]].
%75 citations counted in INSPIRE as of 04 Apr 2020


\bibitem{Weinberg:1990rz}
S.~Weinberg,
%``Nuclear forces from chiral Lagrangians,''
Phys.\ Lett.\ B \textbf{251}, 288-292 (1990)
doi:10.1016/0370-2693(90)90938-3
%1277 citations counted in INSPIRE as of 04 Apr 2020

%\cite{Weinberg:1991um}
\bibitem{Weinberg:1991um}
S.~Weinberg,
%``Effective chiral Lagrangians for nucleon - pion interactions and nuclear forces,''
Nucl.\ Phys.\ B \textbf{363}, 3-18 (1991)
doi:10.1016/0550-3213(91)90231-L
%1211 citations counted in INSPIRE as of 04 Apr 2020

%\cite{Ordonez:1992xp}
\bibitem{Ordonez:1992xp}
C.~Ordonez and U.~van Kolck,
%``Chiral lagrangians and nuclear forces,''
Phys.\ Lett.\ B \textbf{291}, 459-464 (1992)
doi:10.1016/0370-2693(92)91404-W
%293 citations counted in INSPIRE as of 04 Apr 2020

%\cite{Ordonez:1993tn}
\bibitem{Ordonez:1993tn}
C.~Ordonez, L.~Ray and U.~van Kolck,
%``Nucleon-nucleon potential from an effective chiral Lagrangian,''
Phys.\ Rev.\ Lett.\  \textbf{72}, 1982-1985 (1994)
doi:10.1103/PhysRevLett.72.1982
%383 citations counted in INSPIRE as of 04 Apr 2020

%\cite{vanKolck:1994yi}
\bibitem{vanKolck:1994yi}
U.~van Kolck,
%``Few nucleon forces from chiral Lagrangians,''
Phys.\ Rev.\ C \textbf{49}, 2932-2941 (1994)
doi:10.1103/PhysRevC.49.2932
%471 citations counted in INSPIRE as of 04 Apr 2020

%\cite{Ordonez:1995rz}
\bibitem{Ordonez:1995rz}
C.~Ordonez, L.~Ray and U.~van Kolck,
%``The Two nucleon potential from chiral Lagrangians,''
Phys.\ Rev.\ C \textbf{53}, 2086-2105 (1996)
doi:10.1103/PhysRevC.53.2086
[arXiv:hep-ph/9511380 [hep-ph]].
%567 citations counted in INSPIRE as of 04 Apr 2020


%\cite{Epelbaum:2002vt}
\bibitem{Epelbaum:2002vt}
E.~Epelbaum, A.~Nogga, W.~Gloeckle, H.~Kamada, U.~G.~Meissner and H.~Witala,
%``Three nucleon forces from chiral effective field theory,''
Phys.\ Rev.\ C \textbf{66}, 064001 (2002)
doi:10.1103/PhysRevC.66.064001
[arXiv:nucl-th/0208023 [nucl-th]].
%516 citations counted in INSPIRE as of 04 Apr 2020

%\cite{Bernard:2007sp}
\bibitem{Bernard:2007sp}
V.~Bernard, E.~Epelbaum, H.~Krebs and U.~Meissner,
%``Subleading contributions to the chiral three-nucleon force. I. Long-range terms,''
Phys.\ Rev.\ C \textbf{77}, 064004 (2008)
doi:10.1103/PhysRevC.77.064004
[arXiv:0712.1967 [nucl-th]].
%189 citations counted in INSPIRE as of 04 Apr 2020

%\cite{Bernard:2011zr}
\bibitem{Bernard:2011zr}
V.~Bernard, E.~Epelbaum, H.~Krebs and U.~Meissner,
%``Subleading contributions to the chiral three-nucleon force II: Short-range terms and relativistic corrections,''
Phys.\ Rev.\ C \textbf{84}, 054001 (2011)
doi:10.1103/PhysRevC.84.054001
[arXiv:1108.3816 [nucl-th]].
%153 citations counted in INSPIRE as of 04 Apr 2020




%\cite{Epelbaum:2014sza}
\bibitem{Epelbaum:2014sza}
E.~Epelbaum, H.~Krebs and U.~Meissner,
%``Precision nucleon-nucleon potential at fifth order in the chiral expansion,''
Phys.\ Rev.\ Lett.\  \textbf{115}, no.12, 122301 (2015)
doi:10.1103/PhysRevLett.115.122301
[arXiv:1412.4623 [nucl-th]].
%237 citations counted in INSPIRE as of 04 Apr 2020


%\cite{Reinert:2017usi}
\bibitem{Reinert:2017usi}
P.~Reinert, H.~Krebs and E.~Epelbaum,
%``Semilocal momentum-space regularized chiral two-nucleon potentials up to fifth order,''
Eur.\ Phys.\ J.\ A \textbf{54}, no.5, 86 (2018)
doi:10.1140/epja/i2018-12516-4
[arXiv:1711.08821 [nucl-th]].
%84 citations counted in INSPIRE as of 04 Apr 2020

%\cite{Dirac:1949cp}
\bibitem{Dirac:1949cp}
P.~A.~Dirac,
%``Forms of Relativistic Dynamics,''
Rev. Mod. Phys. \textbf{21}, 392-399 (1949)
doi:10.1103/RevModPhys.21.392
%1777 citations counted in INSPIRE as of 24 Apr 2020

%\cite{Girlanda:2010ya}
\bibitem{Girlanda:2010ya}
L.~Girlanda, S.~Pastore, R.~Schiavilla and M.~Viviani,
%``Relativity constraints on the two-nucleon contact interaction,''
Phys. Rev. C \textbf{81}, 034005 (2010)
doi:10.1103/PhysRevC.81.034005
[arXiv:1001.3676 [nucl-th]].
%36 citations counted in INSPIRE as of 24 Apr 2020


%\cite{Manohar:1983md}
\bibitem{Manohar:1983md}
A.~Manohar and H.~Georgi,
%``Chiral Quarks and the Nonrelativistic Quark Model,''
Nucl. Phys. B \textbf{234}, 189-212 (1984)
doi:10.1016/0550-3213(84)90231-1
%2079 citations counted in INSPIRE as of 08 Jul 2020

{
%\cite{Forest:1995sg}
\bibitem{Forest:1995sg}
J.~L.~Forest, V.~R.~Pandharipande and J.~L.~Friar,
%``Relativistic nuclear Hamiltonians,''
Phys. Rev. C \textbf{52}, 568-575 (1995)
doi:10.1103/PhysRevC.52.568
%61 citations counted in INSPIRE as of 17 Sep 2020

%\cite{Krajcik:1974nv}
\bibitem{Krajcik:1974nv}
R.~A.~Krajcik and L.~L.~Foldy,
%``Relativistic center-of-mass variables for composite systems with arbitrary internal interactions,''
Phys. Rev. D \textbf{10}, 1777-1795 (1974)
doi:10.1103/PhysRevD.10.1777
%153 citations counted in INSPIRE as of 18 Sep 2020

%\cite{Friar:1975zza}
\bibitem{Friar:1975zza}
J.~L.~Friar,
%``Relativistic effects on the wave function of a moving system,''
Phys. Rev. C \textbf{12}, 695-698 (1975)
doi:10.1103/PhysRevC.12.695
%68 citations counted in INSPIRE as of 18 Sep 2020
}

  %\cite{Entem:2017gor}
\bibitem{Entem:2017gor}
D.~Entem, R.~Machleidt and Y.~Nosyk,
%``High-quality two-nucleon potentials up to fifth order of the chiral expansion,''
Phys.\ Rev.\ C \textbf{96}, no.2, 024004 (2017)
doi:10.1103/PhysRevC.96.024004
[arXiv:1703.05454 [nucl-th]].
%113 citations counted in INSPIRE as of 04 Apr 2020

%\cite{Piarulli:2014bda}
\bibitem{Piarulli:2014bda}
M.~Piarulli, L.~Girlanda, R.~Schiavilla, R.~Navarro Pérez, J.~Amaro and E.~Ruiz Arriola,
%``Minimally nonlocal nucleon-nucleon potentials with chiral two-pion exchange including $\Delta$ resonances,''
Phys.\ Rev.\ C \textbf{91}, no.2, 024003 (2015)
doi:10.1103/PhysRevC.91.024003
[arXiv:1412.6446 [nucl-th]].
%108 citations counted in INSPIRE as of 04 Apr 2020

  %\cite{Girlanda:2016bze}
\bibitem{Girlanda:2016bze}
L.~Girlanda, A.~Kievsky, M.~Viviani and L.~Marcucci,
%``Progress in the quest for a realistic three-nucleon force,''
PoS \textbf{CD15}, 103 (2016)
doi:10.22323/1.253.0103
%0 citations counted in INSPIRE as of 04 Apr 2020

%\cite{Girlanda:2016neb}
\bibitem{Girlanda:2016neb}
L.~Girlanda, A.~Kievsky, M.~Viviani and L.~E.~Marcucci,
%``Tuning the 3N force from 3N scattering data,''
EPJ Web Conf.\  \textbf{113}, 04009 (2016)
doi:10.1051/epjconf/201611304009
%4 citations counted in INSPIRE as of 04 Apr 2020


%\cite{Girlanda:2018xrw}
\bibitem{Girlanda:2018xrw}
L.~Girlanda, A.~Kievsky, M.~Viviani and L.~Marcucci,
%``Short-range three-nucleon interaction from A=3 data and its hierarchical structure,''
Phys.\ Rev.\ C \textbf{99}, no.5, 054003 (2019)
doi:10.1103/PhysRevC.99.054003
[arXiv:1811.09398 [nucl-th]].
%11 citations counted in INSPIRE as of 04 Apr 2020




\bibitem{Friar:1977xh}
J.~L.~Friar,
%``Pion Exchange Contributions to the Nuclear Charge, Current, and Hamiltonian Operators,''
Annals Phys.\  \textbf{104}, 380-426 (1977)
doi:10.1016/0003-4916(77)90337-2
%139 citations counted in INSPIRE as of 04 Apr 2020
%\cite{Friar:1979by}

\bibitem{Friar:1979by}
J.~L.~Friar,
%``Retardation, quasipotential equations, and relativistic corrections to the deuteron charge operator,''
Phys.\ Rev.\ C \textbf{22}, 796-812 (1980)
doi:10.1103/PhysRevC.22.796
%65 citations counted in INSPIRE as of 04 Apr 2020

\bibitem{Pastore:2011ip}
S.~Pastore, L.~Girlanda, R.~Schiavilla and M.~Viviani,
%``The two-nucleon electromagnetic charge operator in chiral effective field theory ($\chi$EFT) up to one loop,''
Phys.\ Rev.\ C \textbf{84}, 024001 (2011)
doi:10.1103/PhysRevC.84.024001
[arXiv:1106.4539 [nucl-th]].
%65 citations counted in INSPIRE as of 04 Apr 2020


\bibitem{Epelbaum:1998ka}
E.~Epelbaum, W.~Gloeckle and U.~Meissner,
%``Nuclear forces from chiral Lagrangians using the method of unitary transformation. 1. Formalism,''
Nucl.\ Phys.\ A \textbf{637}, 107-134 (1998)
doi:10.1016/S0375-9474(98)00220-6
[arXiv:nucl-th/9801064 [nucl-th]].
%307 citations counted in INSPIRE as of 04 Apr 2020


\bibitem{Epelbaum:1999dj}
E.~Epelbaum, W.~Gloeckle and U.~Meissner,
%``Nuclear forces from chiral Lagrangians using the method of unitary transformation. 2. The two nucleon system,''
Nucl.\ Phys.\ A \textbf{671}, 295-331 (2000)
doi:10.1016/S0375-9474(99)00821-0
[arXiv:nucl-th/9910064 [nucl-th]].
%381 citations counted in INSPIRE as of 04 Apr 2020


\bibitem{Kolling:2011mt}
S.~Kolling, E.~Epelbaum, H.~Krebs and U.~Meissner,
%``Two-nucleon electromagnetic current in chiral effective field theory: One-pion exchange and short-range contributions,''
Phys.\ Rev.\ C \textbf{84}, 054008 (2011)
doi:10.1103/PhysRevC.84.054008
[arXiv:1107.0602 [nucl-th]].
%86 citations counted in INSPIRE as of 04 Apr 2020


\bibitem{Krebs:2016rqz}
H.~Krebs, E.~Epelbaum and U.~Meissner,
%``Nuclear axial current operators to fourth order in chiral effective field theory,''
Annals Phys.\  \textbf{378}, 317-395 (2017)
doi:10.1016/j.aop.2017.01.021
[arXiv:1610.03569 [nucl-th]].
%50 citations counted in INSPIRE as of 04 Apr 2020

  %\cite{Kievsky:1999nw}
\bibitem{Kievsky:1999nw}
A.~Kievsky,
%``Phenomenological spin orbit three-body force,''
Phys.\ Rev.\ C \textbf{60}, 034001 (1999)
doi:10.1103/PhysRevC.60.034001
[arXiv:nucl-th/9905045 [nucl-th]].
%62 citations counted in INSPIRE as of 05 Apr 2020




%\cite{Golak:2014ksa}
\bibitem{Golak:2014ksa}
J.~Golak, R.~Skibinski, K.~Topolnicki, H.~Witala, E.~Epelbaum, H.~Krebs, H.~Kamada, U.~Meissner, V.~Bernard, P.~Maris, J.~Vary, S.~Binder, A.~Calci, K.~Hebeler, J.~Langhammer, R.~Roth, A.~Nogga, S.~Liebig and D.~Minossi,
%``Low-energy neutron-deuteron reactions with N3LO chiral forces,''
Eur. Phys. J. A \textbf{50}, 177 (2014)
doi:10.1140/epja/i2014-14177-7
[arXiv:1410.0756 [nucl-th]].
%38 citations counted in INSPIRE as of 26 Apr 2020


  \end{thebibliography}
\end{document}